\documentclass[prb,twocolumn,amsmath,amssymb]{revtex4}

\usepackage{graphicx}
\usepackage{dcolumn}

\begin{document}

\title{Quantum dynamics of an Ising spin-chain in a random transverse field}

\author{Xun Jia}
\author{Sudip Chakravarty}
\email{sudip@physics.ucla.edu}

\affiliation{Department of Physics and Astronomy, University of
California Los Angeles, Los Angeles, CA 90095-1547}

\date{\today}

\begin{abstract}
    We consider an  Ising spin-chain in a random transverse
    magnetic field and compute the  zero temperature wave vector  and frequency dependent dynamic structure factor numerically
    by using Jordan-Wigner transformation. Two types of
    distributions of magnetic fields are introduced.
    For a rectangular distribution, a dispersing branch is observed,
    and disorder tends to broaden the dispersion
    peak and close the excitation gap. For a binary distribution, a
    non-dispersing branch at almost zero energy is obtained. We discuss the relationship of our work
   to the  neutron scattering measurement in $\mathrm{LiHoF_4}$.
\end{abstract}

\maketitle

Calculation of real-time dynamics of a correlated quantum system
with an infinite number of degrees of freedom are few and far
between. Except for  isolated examples,  construction of real-time
behavior from imaginary-time correlation functions (more amenable to
numerical methods)  by analytic continuation is fraught with various
instabilities. The theoretical challenge is particularly acute
because neutron scattering experiments often provide a rather
detailed map of the frequency, $\omega$, and the wave vector, $\bf
k$, dependent dynamical structure factor, $S({\bf k},\omega)$.

The second motivation comes from the desire to study the dynamics of
a quantum phase transition involving a zero temperature quantum
critical point. In this respect, the Ising spin chain in a
transverse
field\cite{deGennes:1963,Pfeuty:1970,Elliott:1970,Stinchcombe:1973}
constitutes a schema from which much can be learned about quantum
criticality,\cite{Sachdev:1999} both with and without disorder.

The third motivation is to examine how the coherence of the
quasiparticle excitations is modified in the presence of quenched
disorder and is triggered by a recent neutron scattering
experiment\cite{Ronnow:2005} in $\mathrm{LiHoF_{4}}$, which connects
the observed low temperature behavior of $S({\bf k},\omega)$ in
terms of the hyperfine coupling of the electronic spins to a nuclear
spin bath, where the Hamiltonian of the electronic spins is given by
an Ising model in a transverse field. Because the hyperfine
splittings are small, we could imagine that, on the time scale of the
electronic motion, this bath will appear essentially quenched,
modulating the quantum fluctuations characterized by the transverse
field. This is still a bit far from the experimental system, as it
is three-dimensional, and the Ising couplings are long-ranged and
dipolar. Nonetheless, we shall see that there are tantalizing
similarities between our calculated structure factor and the
experimental one in a given direction of the reciprocal space
$(2,0,0) \to (1,0,0)$ (in reciprocal units). A more appropriate
comparison should be with quasi-one dimensional spin systems that
are intentionally disordered.

Finally,  the role of disorder in a quantum critical system is an important
 subject in itself and is certainly not fully understood. In the presence of
disorder, there are rare regions with couplings stronger than the
average, which results in the Griffiths-McCoy
singularities\cite{Griffiths:1969,McCoy:1969}. Although the effect
is weak in a classical system,  it becomes important in quantum
systems, especially in low dimensions\cite{DFisher:1995,Igloi:2005}.

The Ising model in a random transverse field  has been studied  both
analytically and
numerically,\cite{McCoy:1973,Shankar:1987,DFisher:1995,Young:1996,Guo:1996,Young:1997,Sachdev:1997,Derzhko:1997,Igloi:1998,Pich:1998}
and a great number of results of physical importance have been
obtained. However, the dynamical structure factor $S(k, \omega)$ has
not been computed for all $k$ and $\omega$ in the random field
model, although some analytical and numerical results  are available
in pure systems for special values of the wave vector. Here we
compute the dynamic structure factor in the presence of two types of
disorder distributions: a rectangular and a binary distribution.

The one-dimensional lattice Hamiltonian we study is
\begin{equation}\label{hamiltonian}
    H=-J\sum_i\sigma_i^z\sigma_{i+1}^z-\sum_ih_i\sigma_i^x ,
\end{equation}
where the  $\sigma$'s are Pauli matrices, and $J$ is positive and
uniform. We shall choose the energy unit such that $J=1.0$. The
fields $h_i$ are random variables. The first model we study is the
rectangular distribution with a mean $h_{ave}$ and a width $h_w$.
The second model is the binary distribution in which $h_i$ is an independent random variable that takes two values: $h_S$ and $h_L$ with probabilities $p$ and
$(1-p)$, respectively. In particular, we choose the parameter $p$ to
be small such that the chain is almost spatially homogeneous,
$h_S<J<h_L$, to ensure that the system is in the paramagnetic phase.
Intuitively, this distribution seems to capture a crude adiabatic representation of the electronic 
spins coupled to a hyperfine spin bath discussed above, where the larger field is the applied transverse field.

We first compute the time-dependent spin-spin correlation functions
$C(n,t)=\overline{\langle\sigma_i^z(t)\sigma_{i+n}^z\rangle}$ at
temperature $T=0$, where the angular brackets denote the average
over ground state and the overline an average over disorder
configurations. The algorithm for evaluating this quantity is
similar to that described  in Ref.~\onlinecite{Young:1997}, except
that we are doing a real time calculation, which entails computation
over complex variables. We first perform a Jordan-Wigner
transformation\cite{Lieb:1961} and then cast the correlation
function $C(n,t)$ in the form of a Pfaffian, which is calculated
efficiently, as described below; after averaging over disorder
configurations, the dynamical structure factor is,
\begin{equation}\label{Fourier}
    S(k,\omega)=\sum_ne^{-ikn}\int dt e^{-i\omega
    t}C(n,t).
\end{equation}

The pfaffian is  the square root of the determinant of
an antisymmetric matrix. Let $X$ be an $N\times N$
($N$ is even) antisymmetric matrix of the form:
\begin{equation}
    X=\left[%
    \begin{array}{cc}
        A & B \\
        -B^T & C \\
    \end{array}\right],
\end{equation}
where $A$ is a $2\times2$ matrix, and $B$, $C$ are matrices of
appropriate  dimensions. From the identity:
\begin{equation}\begin{split}
    \left[%
    \begin{array}{cc}
        I_2 & 0 \\
        B^TA^{-1} & I_{(N-2)} \\
    \end{array}\right] &X
    \left[%
    \begin{array}{cc}
    I_2 & -A^{-1}B \\
    0 & I_{(N-2)}\\
    \end{array}\right] \\&=
    \left[%
    \begin{array}{cc}
    A & 0 \\
    0 & C+B^TA^{-1}B \\
    \end{array}\right]
\end{split}
\end{equation}
where $I_n$ is the unit matrix of dimension $n$, we have:
\begin{equation}
   \mathrm{Det}(X)=\mathrm{Det}(A)\mathrm{Det}(C+B^TA^{-1}B)
\end{equation}
Since $A$ is also antisymmetric, of the form
\begin{equation}
    A=\left[%
    \begin{array}{cc}
    0 & a_{12} \\
    -a_{12} & 0 \\
    \end{array}\right] ,
\end{equation}
it is easy to invert $A$, and hence calculate $C+B^TA^{-1}B$.
Since the matrix $C+B^TA^{-1}B$ is also antisymmetric of
dimension $(N-2)$, the above procedure can be repeated, and the
determinant of $X$ is given by  the product of $2\times2$ determinants. Finally, since
$\sqrt{\mathrm{Det}(A)}=a_{12}$, the pfaffian of matrix $X$ will be
simply a product of $N/2$ numbers obtained from those $2\times2$
matrices in the above procedure. Because it is not necessary for $A$
to be a $2\times2$ matrix at the upper left corner of $X$, for
the stability of the algorithm, $A$ is chosen to be a diagonal block such that $|\mathrm{Det} (A)|$ is the largest.

The calculations were performed for a lattice of 160 sites of which
32 sites in the middle  were used for evaluating $C(n,t)$. Since we are interested in the gapped phase of the model where the correlation length is finite, the size of the system was found to be sufficient by checking the difference between the results corresponding to various sizes. A window
function, $\exp(-\epsilon_1|t|-\epsilon_2|n|)$, was applied to the
Fourier transform in (\ref{Fourier}) to reduce the cutoff effects in
$S(k,\omega)$.\cite{Derzhko:1997} The chosen small values
$\epsilon_{1}=\epsilon_{2}=0.05$ were found to be sufficient for our
numerical purposes; the range of the time integration used was
$-80\le t \le 80$. Free boundary condition was imposed for
simplicity, and the results were averaged over 100 realizations
of disorder. We explicitly checked that our conclusions are unaffected 
in going from say 20 to 100 realizations of disorder. The reason is that
we are interested in some robust features of the spectra of excitations 
in the quantum paramagnetic phase rather than the critical behavior or 
similar such subtle properties.
\begin{figure}[htb]
\includegraphics[scale=0.6]{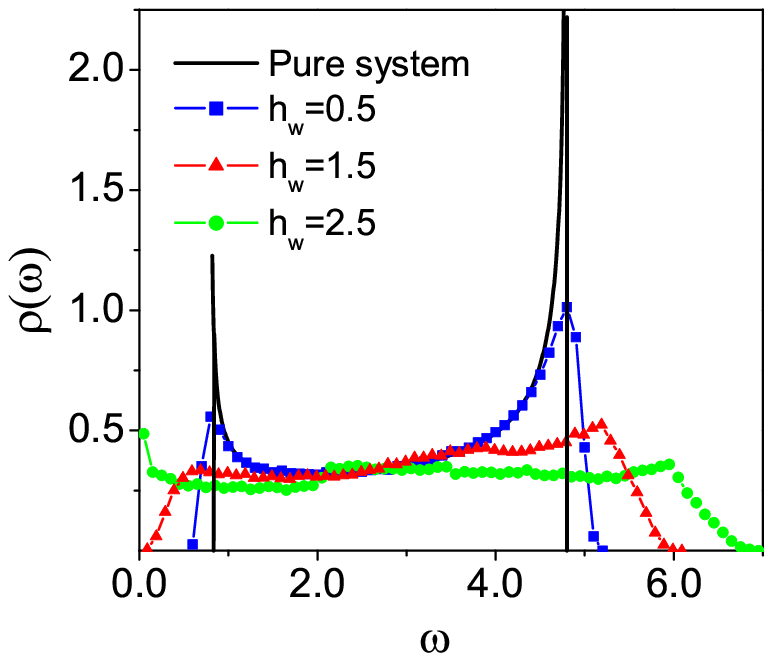}
\caption{(Color online) The density of states of Jordan-Wigner fermions, $\rho(\omega)$, for the
rectangular distribution, for different disorder strengths.
The $\rho(\omega)$ is broadened as $h_w$ increases, and states with
gapless excitations are
available at zero energy only in the high disorder limit.} 
\label{dosrectangular}
\end{figure}

\begin{figure}[htb]
\includegraphics[scale=0.5]{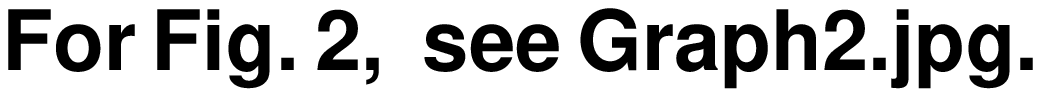}
\caption{(Color online) The dynamic structure factor $S(k,\omega)$:
(a)$\sim$(c) for the rectangular disorder distribution in the
paramagnetic phase: $h_{ave}=1.4$, and $h_w=0.5, 1.5, 2.5$
respectively; (d) $S(k,\omega)$ for pure system with $h=1.4$ and
$J=1.0$. Dashed  line shows the theoretical dispersion relation.}
\label{rectangular}
\end{figure}

To access the paramagnetic
regime, in the rectangular distribution case, not too far away from the quantum critical point, we set
$h_{ave}=1.4$. The density of states for Jordan-Wigner fermions is shown in
Fig.~\ref{dosrectangular}. As we increase disorder, the density of
states (DOS) gets broader,  and only for strong disorder a few states
appear at zero energy, resulting in gapless excitations in this
extreme limit. When the disorder width is small, for example
$h_w=0.5$, and the system is still in the paramagnetic regime
everywhere, there is only a single dispersing branch, and the
excitation remains gapped, as in Fig.~\ref{rectangular}(a). As we
increase $h_w$, the dynamic structure factor can, in principle,  have two
branches because the DOS can acquire states at zero energy. Clearly, the
dispersing branch,  as shown in Fig.~\ref{rectangular}(b),  has very low spectral weight everywhere. In
principle, there could be a strong non-dispersing branch at $\omega=0$.
However, since the excitation gap is closed at this value of
disorder strength, and since both branches are significantly
broadened due to  randomness, the non-dispersing peak may be mixed
with the dispersing branch, and perhaps we are not able to observe
them separately. In the high disorder limit, $h_w=2.5$, almost all
the intensity concentrates at $(k,\omega)=(0,0)$, but there appears be a ghost of the dispersing branch, see Fig.~\ref{rectangular}(c). An important feature
of Figs.~\ref{rectangular}(b) and (c) is the horizontal stripe-like
patterns, which indicate excitation modes that do not disperse,
namely the localized modes. Our code was  checked by comparing with the
exact result for the pure system for which $S(k,\omega)$ has a
single dispersive branch (see below), though the peak is not a delta
function given the finite size of our system; see Fig.~\ref{rectangular}(d).

\begin{figure}[htb]
\includegraphics[scale=0.6]{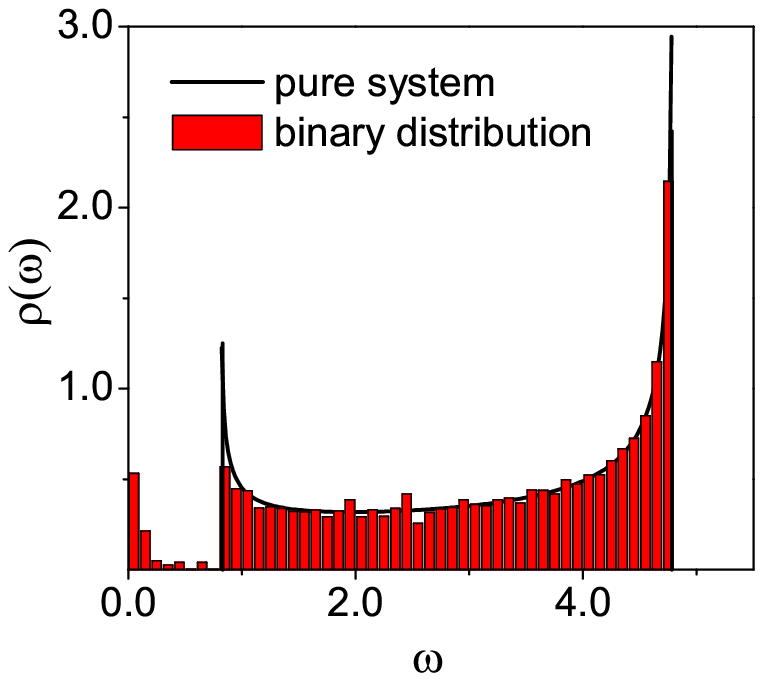}
\caption{(Color online) The density of states $\rho(\omega)$ for the
binary distribution at $h_L=1.4$. Clearly, a few new states are
allowed at zero energy, as the disorder is turned on.} \label{dosbinary}
\end{figure}

\begin{figure}[htb]
\includegraphics[scale=0.5]{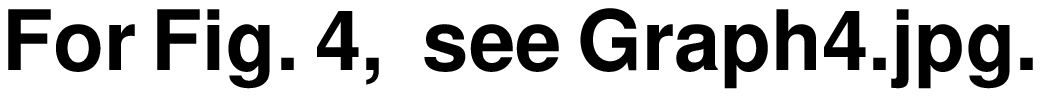}
\caption{(Color online) (a)$\sim$(f) Dynamic structure factor
$S(k,\omega)$ for the binary distribution in the paramagnetic
phase. Parameters are $J=1.0$, $h_S=0.1$, $p=0.05$, and $h_L=1.1,
1.2, \ldots, 1.6$ from (a) through (f). Both dispersing and
non-dispersing branches exist, and the excitation is gapped.}
\label{binary}
\end{figure}

In contrast, the binary distribution is quite remarkable. We first choose $h_L=1.4$ so
that the system is in the paramagnetic regime, and  set $h_S=0.1$,
and $p=0.05$. The density of states in
Fig.~\ref{dosbinary} shows zero energy states separated by a gap from the states at higher energy. The calculated
$S(k,\omega)$ is shown in Figs.~\ref{binary}(a) through (f), where $h_L=1.1,
1.2, \ldots, 1.6$. While the dispersing branch is broadened due to
disorder,  the weight of the central peak around
$(k,\omega)=(0,0)$ is so high that a non-dispersing branch can
extend quite far away from the origin.

\begin{figure}[htb]
\includegraphics[scale=0.8]{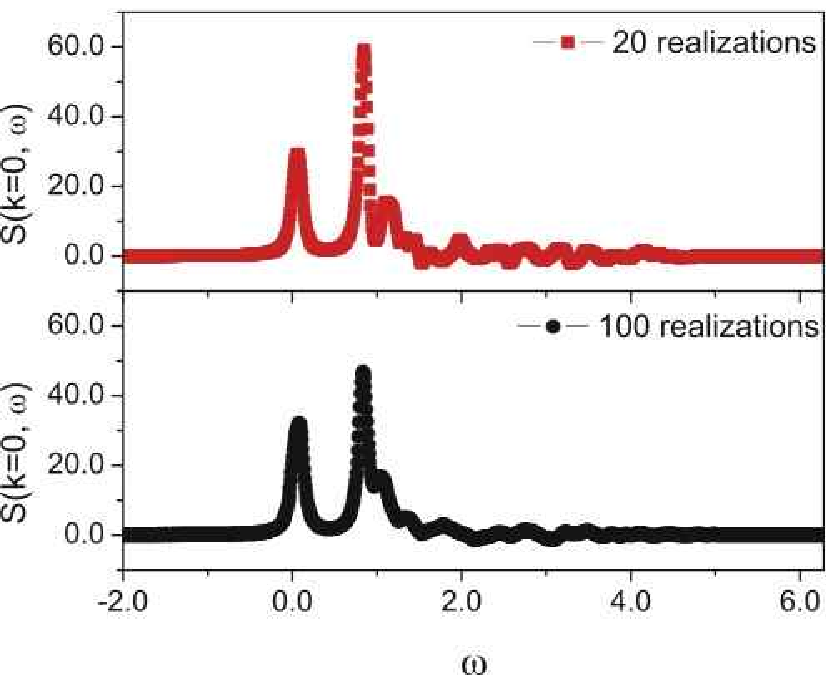}
\caption{(Color online) A cut of $S(0,\omega)$ corresponding to Fig.~\ref{binary}(d) for 20 and 100 realizations of disorder.}
\label{cutk0}
\end{figure}

In Fig.~\ref{cutk0}, we show a cut of $S(k,\omega)$ at $k=0$, to illustrate that there is hardly any difference between the results averaged over 20 realizations and those averaged over 100 realizations.

\begin{figure}[htb]
\includegraphics[scale=0.8]{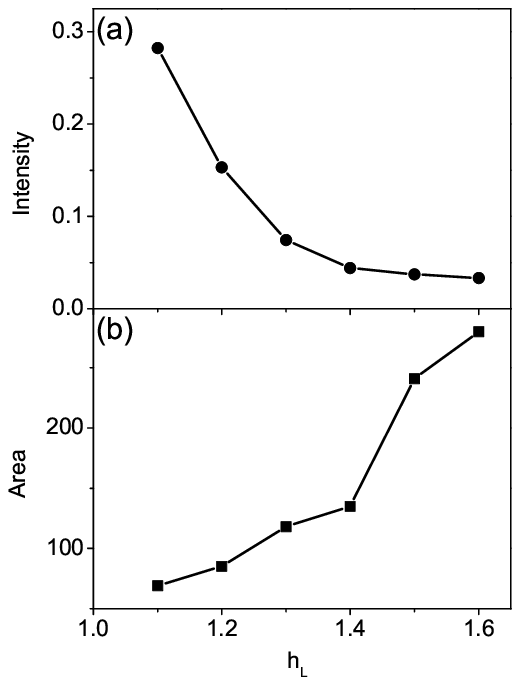}
\caption{The binary distribution: as  the quantum
phase transition is approached from the paramagnetic regime, the central peak
intensity $I_h$ increases monotonically,
and the area $\Delta\Omega$ corresponding to the half the value of the peak decreases.}
\label{peak}
\end{figure}

Let us  define the weight of the central peak as
\begin{equation}\label{intensity}
I_{h}=\frac{\int_{\Delta\Omega}S^2(k,\omega)\mathrm{d}k\mathrm
{d}\omega}{\int_{\Omega}S^2(k,\omega)\mathrm{d}k\mathrm
{d}\omega}
\end{equation}
where $\Omega$ in the denominator is the entire $(k,\omega)$
domain, while $\Delta\Omega$ is defined as follows: we first find
the maximum of the central peak $S_{\textrm{max}}(k,\omega)$, and then define the domain $\Delta\Omega$  such that
$S(k,\omega)>S_{\textrm{max}}(k,\omega)/\sqrt{2}$ for $(k,\omega)\in\Delta\Omega$.
Note that in (\ref{intensity}) we integrate $S^2(k,\omega)$ rather
than $S(k,\omega)$.  This is because small  numerical errors can result in
slightly negative values of $S(k,\omega)$ for some $(k,\omega)$,
which is obviously unphysical. The dependences of $I_h$ and
 $\Delta\Omega$ on $h_L$ are plotted in Fig.~\ref{peak}. As we
tune $h_L$ from $1.6$ to $1.1$, $I_h$ increases monotonically, while
the region $\Delta\Omega$ shrinks. We conclude that,
as  the quantum phase transition is approached, the weight is
transferred from the dispersing branch to the non-dispersing
peak.

In addition to the singularity due to the quantum phase transition,
in disordered systems there is also the Griffiths-McCoy singularity:
the disorder will drive some rare regions into a phase different
from the rest. For the pure Ising chain when $h_i=h$, the
dispersion relation is\cite{Pfeuty:1970}
\begin{equation}\label{dispersion}
    \omega=2\sqrt{J^2+h^2-2Jh\cos{k}}.
\end{equation}
The excitation gap $\Delta=2|h-J|$ occurs at $k=0$. As $h\to J$, the
excitation gap closes at the quantum critical point. However, if
disorder is  strong enough, there is nonzero probability to find
some regions in which the spins are strongly coupled, such that
$h_i<J$ holds, and thus the cluster is ferromagnetic. Calculations,
similar to that given in Ref.~\onlinecite{Sachdev:1999},\cite{footnote} show that these
clusters give rise to the non-dispersing peak at zero energy, as we describe below.

For the binary distribution, the
system is almost homogeneous except for the rare regions of strongly
coupled clusters where $h_{i}=h_{S}$ for all sites inside the
cluster. At shorter length scales, the behavior of the pure system
dominates, and this leads to the dispersing branch. At longer scales
the effects of disorder become important, resulting in the zero
energy peak. The autocorrelation function $S(\omega)$, which is the
integral of $S(k,\omega)$ over $k$ can be approximated in the
following manner. The normalized
probability that a given site belongs to a ferromagnetic cluster of
length $L$-sites is $P(L)=Lp^{L-1}(1-p)^2$. When $h_i=0$ the
two-fold degenerate ground state within a cluster is far away from
the excited states of energy of order $2J$; the perturbation
$h_i=h_S$, will split the ground state by
$\tilde{g}\mathrm{e}^{-cL}$, where $\tilde{g}$ and $c$ are unknown
positive constants determined by the details of the Hamiltonian. The
form of the splitting results from large order in perturbation
theory, however. If we treat these clusters as independent and
average over disorder, or equivalently integrate over the cluster
size $L$, we get
\begin{eqnarray}\label{somega}
    S(\omega)&\sim&\int\mathrm{d}LP(L)\delta(\omega-\tilde{g}\mathrm{e}^{-cL}) \nonumber \\
   &\sim& \frac{(1-p)^2}{p}\frac{\ln(\tilde{g}/\omega)}{\omega}(\frac{\tilde{g}}{\omega})^{\ln(p)/c},
\end{eqnarray}
which diverges at $\omega=0$ if $1> |\ln p|/c$ modulo logarithmic
corrections. We have verified that indeed the divergence disappears
for sufficiently small values of $p$ (the peak of $S(\omega)$ is
then shifted to $\omega$ greater but close to zero, instead), but a
detailed verification of this result appears to be difficult.

It is remarkable that a numerically  exact solution of a simplified one-dimensional model (binary distribution) can capture some of the experimental features in $\mathrm{LiHoF_4}$ and shed interesting light on the role of disorder on the dynamics of a prototypical quantum critical point. It would be of course interesting to experimentally study intentionally disordered systems that are closer to the model discussed here.

We thank Yifei Lou for
suggesting to us the powerful calculational scheme described in the text. This work was
supported by the NSF under the grant DMR-0411931.

\end{document}